\documentstyle[12pt]{article}
\input psfig
\catcode`\@=11
\long\def\@makefntext#1{ 
\protect\noindent \hbox to 3.2pt {\hskip-.9pt
$^{{\ninerm\@thefnmark}}$\hfil}#1\hfill} 

\def\thefootnote{\fnsymbol{footnote}}
 \def\@makefnmark{\hbox to 0pt{$^{\@thefnmark}$\hss}}  

\def\ps@myheadings{\let\@mkboth\@gobbletwo
\def\@oddhead{\hbox{} 
\rightmark\hfil\ninerm\thepage}
\def\@oddfoot{}\def\@evenhead{\ninerm\thepage\hfil 
\leftmark\hbox{}}\def\@evenfoot{}
\def\sectionmark##1{}\def\subsectionmark##1{}}

\catcode`\@=11
\long\def\@makefntext#1{ 
\protect\noindent \hbox to 3.2pt {\hskip-.9pt
$^{{\ninerm\@thefnmark}}$\hfil}#1\hfill} 

\def\thefootnote{\fnsymbol{footnote}}
 \def\@makefnmark{\hbox to 0pt{$^{\@thefnmark}$\hss}}  

\def\ps@myheadings{\let\@mkboth\@gobbletwo
\def\@oddhead{\hbox{} 
\rightmark\hfil\ninerm\thepage}
\def\@oddfoot{}\def\@evenhead{\ninerm\thepage\hfil 
\leftmark\hbox{}}\def\@evenfoot{}
\def\sectionmark##1{}\def\subsectionmark##1{}}

\begin{document}

\newcommand{\symbolfootnote}{\renewcommand{\thefootnote}
	{\fnsymbol{footnote}}}
\renewcommand{\thefootnote}{\fnsymbol{footnote}}
\newcommand{\alphfootnote}
	{\setcounter{footnote}{0}
	 \renewcommand{\thefootnote}{\sevenrm\alph{footnote}}}

\newcounter{sectionc}\newcounter{subsectionc}\newcounter{subsubsectionc}
\renewcommand{\section}[1] {\vspace{0.6cm}\addtocounter{sectionc}{1}
\setcounter{subsectionc}{0}\setcounter{subsubsectionc}{0}\noindent
	{\bf\thesectionc. #1}\par\vspace{0.4cm}}
\renewcommand{\subsection}[1] {\vspace{0.6cm}\addtocounter{subsectionc}{1}
	\setcounter{subsubsectionc}{0}\noindent
	{\it\thesectionc.\thesubsectionc. #1}\par\vspace{0.4cm}}
\renewcommand{\subsubsection}[1]
{\vspace{0.6cm}\addtocounter{subsubsectionc}{1}
	\noindent {\rm\thesectionc.\thesubsectionc.\thesubsubsectionc.
	#1}\par\vspace{0.4cm}}
\newcommand{\nonumsection}[1] {\vspace{0.6cm}\noindent{\bf #1}
	\par\vspace{0.4cm}}
\newcounter{appendixc}
\newcounter{subappendixc}[appendixc]
\newcounter{subsubappendixc}[subappendixc]
\renewcommand{\thesubappendixc}{\Alph{appendixc}.\arabic{subappendixc}}
\renewcommand{\thesubsubappendixc}
	{\Alph{appendixc}.\arabic{subappendixc}.\arabic{subsubappendixc}}

\renewcommand{\appendix}[1] {\vspace{0.6cm}
        \refstepcounter{appendixc}
        \setcounter{figure}{0}
        \setcounter{table}{0}
        \setcounter{equation}{0}
        \renewcommand{\thefigure}{\Alph{appendixc}.\arabic{figure}}
        \renewcommand{\thetable}{\Alph{appendixc}.\arabic{table}}
        \renewcommand{\theappendixc}{\Alph{appendixc}}
        \renewcommand{\theequation}{\Alph{appendixc}.\arabic{equation}}
        \noindent{\bf Appendix \theappendixc #1}\par\vspace{0.4cm}}
\newcommand{\subappendix}[1] {\vspace{0.6cm}
        \refstepcounter{subappendixc}
        \noindent{\bf Appendix \thesubappendixc. #1}\par\vspace{0.4cm}}
\newcommand{\subsubappendix}[1] {\vspace{0.6cm}
        \refstepcounter{subsubappendixc}
        \noindent{\it Appendix \thesubsubappendixc. #1}
	\par\vspace{0.4cm}}

\def\abstracts#1{{
	\centering{\begin{minipage}{30pc}\tenrm\baselineskip=12pt\noindent
	\centerline{\tenrm ABSTRACT}\vspace{0.3cm}
	\parindent=0pt #1
	\end{minipage} }\par}}

\newcommand{\bibit}{\it}
\newcommand{\bibbf}{\bf}
\renewenvironment{thebibliography}[1]
	{\begin{list}{\arabic{enumi}.}
	{\usecounter{enumi}\setlength{\parsep}{0pt}
\setlength{\leftmargin 1.25cm}{\rightmargin 0pt}
	 \setlength{\itemsep}{0pt} \settowidth
	{\labelwidth}{#1.}\sloppy}}{\end{list}}

\topsep=0in\parsep=0in\itemsep=0in
\parindent=1.5pc

\newcounter{itemlistc}
\newcounter{romanlistc}
\newcounter{alphlistc}
\newcounter{arabiclistc}
\newenvironment{itemlist}
    	{\setcounter{itemlistc}{0}
	 \begin{list}{$\bullet$}
	{\usecounter{itemlistc}
	 \setlength{\parsep}{0pt}
	 \setlength{\itemsep}{0pt}}}{\end{list}}

\newenvironment{romanlist}
	{\setcounter{romanlistc}{0}
	 \begin{list}{$($\roman{romanlistc}$)$}
	{\usecounter{romanlistc}
	 \setlength{\parsep}{0pt}
	 \setlength{\itemsep}{0pt}}}{\end{list}}

\newenvironment{alphlist}
	{\setcounter{alphlistc}{0}
	 \begin{list}{$($\alph{alphlistc}$)$}
	{\usecounter{alphlistc}
	 \setlength{\parsep}{0pt}
	 \setlength{\itemsep}{0pt}}}{\end{list}}

\newenvironment{arabiclist}
	{\setcounter{arabiclistc}{0}
	 \begin{list}{\arabic{arabiclistc}}
	{\usecounter{arabiclistc}
	 \setlength{\parsep}{0pt}
	 \setlength{\itemsep}{0pt}}}{\end{list}}

\newcommand{\fcaption}[1]{
        \refstepcounter{figure}
        \setbox\@tempboxa = \hbox{\tenrm Fig.~\thefigure. #1}
        \ifdim \wd\@tempboxa > 6in
           {\begin{center}
        \parbox{6in}{\tenrm\baselineskip=12pt Fig.~\thefigure. #1 }
            \end{center}}
        \else
             {\begin{center}
             {\tenrm Fig.~\thefigure. #1}
              \end{center}}
        \fi}

\newcommand{\tcaption}[1]{
        \refstepcounter{table}
        \setbox\@tempboxa = \hbox{\tenrm Table~\thetable. #1}
        \ifdim \wd\@tempboxa > 6in
           {\begin{center}
        \parbox{6in}{\tenrm\baselineskip=12pt Table~\thetable. #1 }
            \end{center}}
        \else
             {\begin{center}
             {\tenrm Table~\thetable. #1}
              \end{center}}
        \fi}

\def\@citex[#1]#2{\if@filesw\immediate\write\@auxout
	{\string\citation{#2}}\fi
\def\@citea{}\@cite{\@for\@citeb:=#2\do
	{\@citea\def\@citea{,}\@ifundefined
	{b@\@citeb}{{\bf ?}\@warning
	{Citation `\@citeb' on page \thepage \space undefined}}
	{\csname b@\@citeb\endcsname}}}{#1}}

\newif\if@cghi
\def\cite{\@cghitrue\@ifnextchar [{\@tempswatrue
	\@citex}{\@tempswafalse\@citex[]}}
\def\citelow{\@cghifalse\@ifnextchar [{\@tempswatrue
	\@citex}{\@tempswafalse\@citex[]}}
\def\@cite#1#2{{$\null^{#1}$\if@tempswa\typeout
	{IJCGA warning: optional citation argument
	ignored: `#2'} \fi}}
\newcommand{\citeup}{\cite}

\def\fnm#1{$^{\mbox{\scriptsize #1}}$}
\def\fnt#1#2{\footnotetext{\kern-.3em
	{$^{\mbox{\sevenrm #1}}$}{#2}}}


\font\twelvebf=cmbx10 scaled\magstep 1
\font\twelverm=cmr10 scaled\magstep 1
\font\twelveit=cmti10 scaled\magstep 1
\font\elevenbf=cmbx10 scaled\magstephalf
\font\elevenrm=cmr10 scaled\magstephalf
\font\tenbf=cmbx10
\font\tenrm=cmr10
\font\tenit=cmti10
\font\ninebf=cmbx9
\font\ninerm=cmr9
\font\nineit=cmti9
\font\eightbf=cmbx8
\font\eightrm=cmr8
\font\eightit=cmti8

\newcommand{\Msun}{M_{\odot}\ }
\newcommand{\Lsun}{L_{\odot}\ }
\newcommand{\hMpc}{{{$h^{-1}$}Mpc}\ }
\newcommand{\Mpc}{\mbox {Mpc}}
\newcommand{\ea}{{et\thinspace al.}\ }

\centerline{\tenbf MIXED COLD-HOT DARK MATTER MODELS}
\smallskip
\centerline{\tenbf WITH SEVERAL MASSIVE NEUTRINO TYPES}
\vspace{0.8cm}

\centerline{\tenrm Dmitri Pogosyan}
\baselineskip=13pt
\centerline{\tenit CITA, University of Toronto}
\baselineskip=12pt
\centerline{\tenit Toronto ONT M5S 1A7, Canada}
\smallskip
\centerline{\tenrm and}
\smallskip
\centerline{\tenrm Alexei Starobinsky}
\baselineskip=13pt
\centerline{\tenit Landau Institute for Theoretical Physics}
\baselineskip=12pt
\centerline{\tenit Kosygina St. 2, Moscow 117334, Russia}

\vspace{0.4cm}
\centerline{\tenit Presented at the 11th Potsdam Workshop on Relativistic
Astrophysics, September 1994}
\centerline{\tenit To be published in Astrophys. J.}
\vspace{0.5cm}
\abstracts{
Mixed cold-hot dark matter cosmological models (CHDM)
with $\Omega_{tot}=1$, approximately flat initial
spectrum of adiabatic perturbations and 1, 2 or 3 types of massive
neutrinos are compared and tested using recent observational data.
The models with 2 or 3 neutrino types of equal mass
permit as the best fit larger values
of both the Hubble constant ($H_0\le 60$ for 2 types,
$H_0\le 65$ for 3 types) and the total $\Omega_{\nu}$
(up to 0.3 for 3 types) than the model with 1 massive
type. Also, they have less problems with abundances of early compact
objects including $Ly-\alpha$ clouds.}

\vfil
\rm\baselineskip=14pt

\section{Introduction}

If the classification of different cosmological models
by a number of additional fundamental parameters used in them to explain
all observational data \cite{s1993,ps2} is applied to inflationary models,
then the model of the first level having only one
fundamental parameter -- an amplitude of perturbations --
appears to be the CDM model with the approximately flat
(Harrison-Zeldovich, or $n\approx 1$) spectrum of initial adiabatic
perturbations. Because of theoretical considerations and observational
uncertainties, it is better to include ``weakly-tilted'' models with
$|n-1|\le 0.1$ into this class, too.

At present, it is clear already that predictions of this model,
though being
not far from  observational data (that is remarkable for a
such a simple model with only one free parameter),
still definitely do not agree with all of them.
Namely, if the free parameter is chosen to fit the data on scales
$(100-1000)h_{50}^{-1}$ Mpc, discrepancy of about twice in perturbation
amplitude arises on scales $(1-10)h_{50}^{-1}$ Mpc,
and vice versa ($h_{50}=H_0/50$, where $H_0$ is the Hubble constant
in km/s/Mpc). Thus, models of
the next (second) level having one more additional constant have
to be considered. Among the best of such models is certainly the mixed
cold+hot dark matter model (CHDM) \cite{shafi84},
for recent analysis see our previous papers
(hereafter PS1 \cite{ps1} and PS2 \cite{ps2}),
as well as \cite{lily93,kly93} and references therein.
In this model, the hot component is
assumed to be the most massive of 3 neutrino species (presumably,
$\tau$-neutrino) with the standard concentration following from
the textbook Big Bang theory, while masses of the other two types of
neutrinos are supposed to be much less and, therefore, unimportant for
cosmology. Then the only new fundamental parameter
is the neutrino mass $m_{\nu}$.

Still the CHDM model with $n\approx 1$ is not without difficulties.
The main of them is connected with later galaxy and quasar formation
in this model as compared to the SCDM model. As a result, only a small
region in the $H_0-\Omega_{\nu}$ plane remains permitted (PS1 \cite{ps1},
some authors have a more pessimistic view \cite{cen94}). The analysis of the
possibility of non-flat (but power law) initial spectra (the article
\cite{lyli94} for
$n < 1$, PS2 \cite{ps2} for detailed analysis of $n \ge 1$ case)
leads to the conclusion that the CHDM works the best with nearly flat
$n\approx (0.95-0.97)$ initial spectra,
which follow from the simplest inflationary
models. Therefore, allowing for this degree of freedom does not improve the fit
to observational data.

In this paper we consider how the cosmological predictions of the CHDM
change if not only one but two or even all three types of neutrino
are massive and contribute to the present density of hot matter. The
total energy density (in terms of critical one) of hot component in this
general case is then $\sum m_{\nu _i}=
23.4\Omega_{\nu}h_{50}^2~{\rm eV}$ for $T_{\gamma}=2.73$K.

This version of CHDM, of course, also corresponds to the next level
of complexity in our classification until the mass ratio of different neutrino
will be either confirmed in laboratory experiments, or
theoretically derived from some underlying theory.
We shall not cover all possible combination of the neutrino masses,
noticing that the case of three equal masses $m _{\nu _1} = m _{\nu _2} =
m _{\nu _3} $ is the one mostly different from a standard model with one
massive neutrino. The predictions for all other sets of $m _{\nu _i}$
(with the same total $\Omega _{\nu}$) lie ``in between'' these two models.
There is no experimental evidence for all three types of neutrino to have
comparable mass. Thus the model with $ m _{\nu _1} = m _{\nu _2} =
m _{\nu _3} $ is mostly interesting as a limiting case. More realistic
is the assumption of two type of neutrino to have mass in cosmologically
interesting range while the rest one (electron neutrino) has a much smaller
mass. Therefore we restrict ourselves with two models
$m _{\nu _1} = 0 ,~ m _{\nu _2} = m _{\nu _3} $ and
$m _{\nu _1} = m _{\nu _2} = m _{\nu _3} $ and compare them with
the original CHDM
with one massive neutrino. We denote by $N_{\nu}$
the number of (equally) massive types of neutrino which mass is now
$m_{\nu}=23.4\Omega_{\nu}/ N_{\nu} h_{50}^2~{\rm eV} $.

\section{CHDM Linear Perturbation Spectrum}

The transfer function $ C(k) $ for the CHDM with one massive
neutrino \mbox{($N_{\nu}=1$)} was determined in our previous
paper \cite{ps1} numerically by solving a system of the Einstein-Vlasov
equations for the evolution of adiabatic perturbations with
a neutrino component treated kinetically and a CDM component as dust.
The correspondent fitting formula is given in \cite{ps2}.
Results of similar computations for the case $N_{\nu}=2, \, 3$ are given
in Fig.1 in comparison with $N_{\nu}=1$ curve for a specific choice
$\Omega _{\nu} = 0.2 $.

\begin{figure}[h]
\centering
\caption{\tenrm\baselineskip=12pt
Upper-left panel: $C(k)$ transfer function for $\Omega_{\nu}=0.2$ and
$N_{\nu}=1,2,3$ massive neutrino types.
Other panels:
Restrictions in the $ H_0 - \Omega_{\nu} $ parameter plane
for $N_{\nu}=1 $ (upper right),  $N_{\nu}=2 $ (lower right) and
$N_{\nu}=3 $ (lower left panel), following from:
a)~fit to Stromlo-APM counts-in-cells. Solid lines correspond to
$\chi ^2 = 2,~7 $ contours;
b)~the $\sigma_8$ condition.
The values $ \sigma_8<1,~0.67$ for $Q_{rms-PS} = 14.3$ $\mu$K
are achieved left to the dashed
(correspondingly rightmost and leftmost) lines. The middle dashed line
corresponds to $\sigma _8 = 1 $ if $Q_{rms-PS} = 19 \,\mu$K;
c)~the combination of the $\sigma_8<0.67$ condition with
quasar and galaxy formation conditions (dotted lines).
The allowed region lies below the upper dotted line
if the limitation set for objects of the mass
$ M = 10^{11} \, {\rm{M}}_{\odot} $ is used
and below the lower one if $ M = 10^{12} \, {\rm{M}}_{\odot} $.
}
\end{figure}

The main difference between models comes from the fact that the same total
mass density fraction $\Omega _{\nu}$ is achieved with different masses of
neutrino. During the matter
dominated stage the presence of hot component distributed at the
beginning of this stage uniformly
due to free-streaming lead to the retarded growth of perturbation in total
density on modes $k R_{nr} < 1 $. Here $ R_{nr} $ is the comoving size of
horizon at the moment when neutrino become nonrelativistic. For the mode $k$
this period continues until $k$ becomes smaller that the
effective neutrino Jeans scale
$k^2_{\rm J}=4\pi G\rho_{tot}\big/({\rm d}p_{\nu}/ {\rm d}\rho_{\nu})
\propto m_{\nu}^{2} t^{-2/3} $ when perturbation in neutrino density
reach the magnitude of inhomogeneities in cold component
(see discussion in PS1 \cite{ps1}).
This period is larger for smaller mass, and
as the result, the amplitude of perturbation at the present moment is
decreased even further from SCDM value. This explains why the transfer
function $ C_{\rm{CHDM}} $ is $\approx 20\% $ lower for larger $N_{\nu}$ at the
intermediate scales
$R_{nr}^{-1} < k < k_{\rm J}(t_0) \approx 11\Omega_{\nu}h{^3_{50}}\,{\rm
Mpc}^{-1}$.
For wavelength shorter that the present Jeans scale $ k > k_{\rm J}(t_0) $
the perturbation in neutrino density did not start to grow up to the present
moment and no additional dependence on $m_{\nu}$ appear. In this limit,
as Fig.1 demonstrates, the resulting amplitude is approximately the
same as far as
the total amount of neutrino $\Omega_{\nu}$ is the same.

\section{Confrontation with Observational Tests}

In Fig.1 we present the restrictions in the $\Omega _{\nu} - H_0 $
parameter plane which follow from several observational tests shown
to be the most illustrative in our previous papers. These tests are:
a) value of the total {\it rms} mass fluctuation $\sigma_8$
at $R=16h{^{-1}_{50}}{\rm{Mpc}}$ ($\sigma_8 <0.67$ based on cluster
abundance data \cite{wef93}) as follows from
COBE measurement of DMR anisotropy $(\Delta T / T) $ which we adopt as
$Q_{rms-PS} = (17.4 \pm 3.1)$ $\mu$K \cite{wri93};
b)~The Stromlo-APM counts in cells \cite{lov92} which limits the
slope of spectrum over the range $(20 - 150) \, h_{50}^{-1}$ Mpc;
c) density of quasars at high redshifts $z \approx 4 $ \cite{er88,haeh93}.
Necessity to produce sufficient
number of quasars in the model puts a lower limit on mass fluctuations
at $\propto 1 $Mpc scale. We follow \cite{haeh93} where the quasar density
criterion at $z=4$ was formulated as a restriction on a mass fraction in bound
objects with mass larger $10^{11} \, {\rm{M}}_{\odot}$:
$ f( \ge 10^{11} \, {\rm{M}}_{\odot} ) \ge 10^{-4} $. Similar estimate
\cite{haeh93} for a fraction of
mass in large galaxies gives $ f( \ge 10^{12} \, {\rm{M}}_{\odot} ) \ge
10^{-5} $ at the same redshift $z=4$. We put greater weight on this second,
more restrictive limit, because, remarkably, it gives similar restrictions
on the CHDM model as the analysis of the abundance of damped $Ly-\alpha$
absorption systems in high-redshift quasar spectra \cite{kly94,ma94}.

The upper-right panel of Fig.1 presents the $\Omega _{\nu} - H_0 $
parameter plane with observational restrictions for standard CHDM with
$n=0.95 $. This plot was discussed in detail both in PS1 \cite{ps1}
and PS2 \cite{ps2}, showing that CHDM
model parameters are restricted to the narrow range of
a low Hubble constant $ H_0 \le 55 $ km/s/Mpc and
the neutrino fraction $\Omega_{\nu} = 0.17-0.28 $ for $H_0 = 50 $.
The first of these limits reflects a problem with unavoidable
high mass fluctuations at the $16 h_{50}^{-1}$ Mpc scale (note that to
set this limit we use a {\it lower} bound on $(\Delta T / T) $,
$Q_{rms-PS} = 14.3$ $\mu$K although recent analysis
suggest higher value $Q_{rms-PS} \approx 20$ $\mu$K),
as well as a wrong shape of the perturbation spectrum over the $l=(20-150)
h_{50}^{-1}$ Mpc interval if $H_0$
is high. The upper bound on $\Omega_{\nu}$ comes from
the combined quasar and $\sigma _8$ conditions
that implies that the slope of the spectrum in the scale range
$(0.7 h_{50}^{-2/3} - 16 h_{50}^{-1})$ Mpc cannot be too steep.

The bottom panels of Fig.1 correspond to two- and three- massive
neutrino models $N_{\nu}=2,~3$. The additional decrease of the transfer
function
in the range $0.01 {\rm Mpc}^{-1} < k < 1 {\rm Mpc}^{-1} $ while keeping
both larger and smaller scales unchanged allow us to noticeably relax
the restrictions on the CHDM models. The major change is the possibility
to accommodate the higher values of Hubble constant. Now both  the shape
of spectrum better fits count-in-cells and the $ \sigma_8 $ limit,
understandably, becomes less severe. For an extreme example $N_{\nu}=3$
we may have $H_0$ as high as $70-75$ km/s/Mpc. For $N_{\nu}=2$ the tests
can be satisfied if $ H_0 \le 65 $ km/s/Mpc, although more comfortable
fit to data is achieved for $ H_0 \le 60 $ ($N_{\nu}=2$) and $ H_0 \le 65 $
($N_{\nu}=3$). Let us stress, however, that
CHDM by itself does not address the problem of the age of the Universe,
which remains the major theoretical objection to $H_0\ge 65$ km/s/Mpc.

Additionally a somewhat higher values of $\Omega_{\nu} $ are now allowed.
Although for $H_0 =50$ the boundaries on total neutrino fraction
remain unchanged $\Omega_{\nu} = 0.15-0.25 $, for $H=60$ and
$N_{\nu}=2,~3$ we have from Fig.1 $ 0.2 < \Omega_{\nu} < 0.3 $. This change
is rather significant as a limit on the sum of neutrino
masses: $3.5 {\rm eV} < \sum m_{\nu _i} < 6 {\rm eV} $
if we take $H_0=50 $ and
$6.7 {\rm eV} < \sum m_{\nu _i} < 10 {\rm eV} $ for $H_0=60$.
On the other hand, the
mass of a single neutrino type is approximately the same
$m_{\nu _i} \simeq 2-5 $ eV.

We can conclude that having several types of neutrino with
cosmologically significant mass of few electron-volts expands the boundary
of parameter space for CHDM model. The allowed region of
$\Omega _{\nu} - H_0$ plane now include not only the small area of
$H_0 \approx 50$, $\Omega _{\nu} \approx 0.2 $ but extends to
$H_0 \approx 60$, $\Omega _{\nu} \approx 0.25 $ for the most physically
interesting version with two massive neutrino, and as far as
$H_0 \approx 70$, $\Omega _{\nu} \approx 0.3 $ for the extreme case
of three equally massive types. Similar results for the $N_{\nu}=2$
case were recently presented in \cite{pr94}.

A.S. was supported in part by the Russian Foundation for
Basic Research, Project Code 93-02-3631, and by Russian Research
Project ``Cosmomicrophysics''. D.P. is grateful for the Organizing Committee
of the 11th Potsdam Workshop on Relativistic Astrophysics for
financial support.

\clearpage

\end{document}